\documentclass[twocolumn,prb,showpacs]{revtex4}
\usepackage{graphicx}
\usepackage{bm}

\begin{document}
\title{Lattice gas description of pyrochlore and checkerboard
antiferromagnets in a strong magnetic field}

\author{M. E. Zhitomirsky}
\affiliation{
Commissariat \`a l'Energie Atomique, DSM/DRFMC/SPSMS, 38054 Grenoble, 
France}
\author{Hirokazu Tsunetsugu}
\affiliation{
Institute for Solid State Physics, University of Tokyo, Kashiwa,
Chiba 277-8581, Japan}

\date{08 March, 2007}

\begin{abstract}
Quantum Heisenberg antiferromagnets on pyrochlore and checkerboard lattices 
in a strong external magnetic field are mapped onto hard-core lattice gases 
with an extended exclusion region. 
The effective models are studied by the exchange Monte Carlo simulations
and by the transfer matrix method.
The transition point and the critical exponents are obtained numerically 
for a square-lattice gas of particles with the second-neighbor exclusion, 
which describes a checkerboard antiferromagnet.
The exact structure of the magnon crystal state is determined for a pyrochlore
antiferromagnet. 
\end{abstract}

\maketitle

\section{Introduction}

A characteristic feature of geometrically frustrated spin models 
is an infinite degeneracy of the classical ground state. Such a 
degeneracy may be lifted by quantum or thermal fluctuations via
the `order from disorder' effect. \cite{villain,shender,henley89}
A closely related property is emergence of small energy scales,
which are described by the effective Hamiltonians acting in the subspace 
of nearly degenerate low-energy quantum states. To date, several types of 
effective models have been developed for frustrated 
antiferromagnets in different regimes.  The first example is 
a celebrated quantum dimer model introduced phenomenologically
by Rokhsar and Kivelson to describe a magnetically disordered 
resontaing valence bond phase of $S=1/2$ antiferromagnets. \cite{RK}
Recently, effective quantum dimer models have been derived
for several realistic spin Hamiltonians in zero and in 
a finite magnetic field. \cite{cabra,mzh05,bergman}
The second example is the effective Hamiltonian for the collinear
states, which applies to  a semiclassical ($S\gg 1$) pyrochlore
antiferromagnet in zero magnetic field. \cite{hizi} The third 
type of effective models describes geometrically frustrated 
magnets at high magnetic fields. \cite{hard1,derzhko04,hard2,richter06} 
These are the so called lattice-gas models considered further 
in the present work.

Previously, the effective lattice-gas models have been derived and 
discussed for the sawtooth chain and a kagom\'e antiferromagnet. 
\cite{hard1,derzhko04,hard2} These two quantum antiferromagnets 
are mapped onto lattice gases of hard-core classical particles 
with the nearest-neighbor exclusion on a one-dimensional chain 
and a triangular lattice, which are 
respectively called the hard-dimer and the hard-hexagon model.
In both cases, the thermodynamics of the effective models has been 
calculated exactly. \cite{baxter}
In the present article, we shall focus on the high-field behavior 
of quantum antiferromagnets on two- and three-dimensional 
pyrochlore lattices. The corresponding effective models do not allow 
exact solution for their thermodynamic properties, therefore, 
we shall study them via a combination of numerical techniques.

\begin{figure}
\begin{center}
\includegraphics[width=0.95\columnwidth]{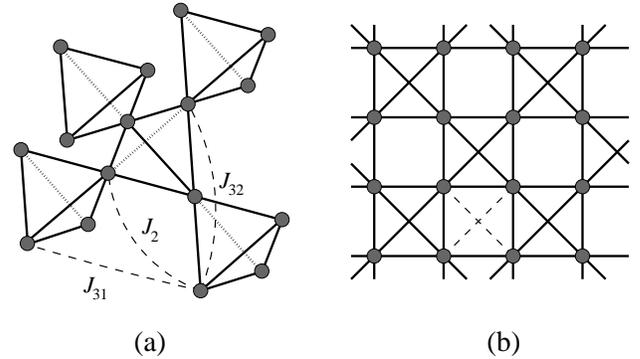}
\end{center}
\caption{ 
Pyrochlore lattice of corner-sharing tetrahedra (a) 
and its two-dimensional analog,
checkerboard lattice (b). Solid lines show
the nearest-neighbor bonds of strength $J$.
Dashed lines indicate possible further neighbor exchanges.
\label{lattices}} 
\end{figure}

The pyrochlore and the checkerboard lattices (Fig.~\ref{lattices})
consist of 3D and 2D networks of corner-sharing tetrahedra
(four-spin blocks).
For the most part of the present work, we disregard weak further neighbor
exchanges and consider the nearest-neighbor Heisenberg model
with an arbitrary spin $S$ in an external magnetic field:
\begin{equation}
\hat{\cal H} = J \sum_{\langle ij\rangle} 
{\bf S}_i\cdot {\bf S}_j
- {\bf H}\cdot \sum_i {\bf S}_i \ .
\label{Hamiltonian}
\end{equation}
Above the saturation field $H_s$ the ground state
of a Heisenberg antiferromagnet is a fully polarized vacuum
$|0\rangle = |\uparrow\uparrow\uparrow...\rangle$,
where all spins are in a state with the maximum possible value
of $S_i^z =S$.
The low-lying excitations are single spin flips
$|i\rangle = S_i^-|0\rangle$, whose dispersion can be straightforwardly
calculated.
The characteristic feature of many geometrically frustrated models
including the two considered models is
presence of the low-energy branch of localized magnons
with the energy $\varepsilon_0 = H-H_s$. 
\cite{schnack,schulenburg,schmidt,mzh03,richter}
The details for the two discussed models are presented
in the subsequent sections.

Localized nature of magnons from the lowest branch allows to
construct a subclass of exact multimagnon states: \cite{schulenburg}
a quantum state with  $m$ localized magnons ($m$-LMs)
occupying spatially separated regions of the lattice, which
are not directly connected by the exchange bonds, is an exact eigenstate
of the quantum Hamiltonian (\ref{Hamiltonian}) and its energy
is given by $m\varepsilon_0$. It has been proven exactly that
isolated LM states correspond
to the lowest energy states in every magnetization subsector. 
\cite{schmidt,hard2} 
In addition, the renormalization arguments suggest 
that delocalized (propagating) states are separated by a finite gap from 
LMs.\cite{hard1,hard2}
Hence, at low temperatures the partition function
of a quantum frustrated antiferromagnet in the vicinity
of the saturation field is determined entirely by LMs:
\begin{equation}
{\cal Z} = \sum_{m=0}^{N_{\rm max}} w(m,N)\,e^{m\mu/T}  \ ,\ \ \
\mu = H_s-H \ ,
\label{Z}
\end{equation}
with $z=e^{\mu/T}$ being the activity. Combinatorial factors 
$w(m,N)$ denote a number of linearly independent $m$-LM states,
while $N_{\rm max}$ is the maximal number of LMs
for a lattice with $N$ sites ($N\rightarrow\infty$).
Mapping to a lattice gas of hard-core objects is used as 
an {\it approximate} way to calculate $w(m,N)$ and $\cal Z$.
The general consequence of Eq.~(\ref{Z}) is the following 
scaling of the total entropy: ${\cal S}=f[(H-H_s)/T]$. 
At $H=H_s$ the entropy is temperature independent and, consequently,
even at $T=0$ a frustrated quantum antiferromagnet has
a finite macroscopic entropy, which is determined purely by the lattice
geometry. The quantum order from disorder mechanism becomes ineffective
for this special value of applied field due to the localized
nature of (exact) quantum states with zero energy.
In Secs.~II and III we describe derivation of lattice gas mapping 
and obtain  conclusions separately for the checkerboard and the pyrochlore 
antiferromagnet. Role of  extra exchanges beyond the nearest-neighbor 
pairs, see Fig.~\ref{lattices}, is briefly discussed in Sec.~IV.

\section{Checkerboard antiferromagnet}

\subsection{Localized magnon states}

\begin{figure}
\begin{center}
\includegraphics[width=0.9\columnwidth]{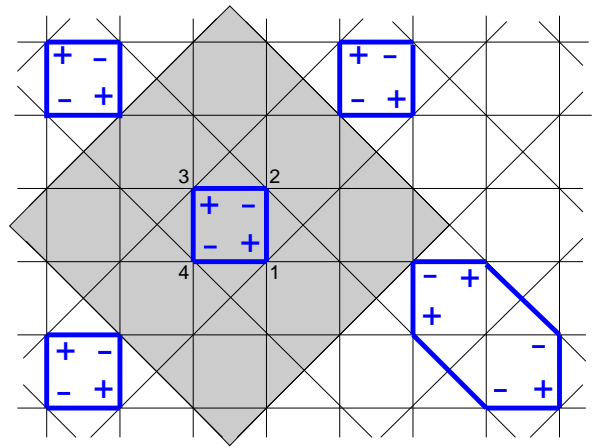}
\end{center}
\caption{(color online). Localized magnons on a checkerboard lattice.
Thick lines indicate positions of one-magnon states with corresponding
phases shown by $+$ and $-$.
The shaded area shows excluded volume around a central square-void
state, which has to be respected for construction of multiparticle
states. Four LMs in the upper-left part satisfy a close-packed condition
(see the text).
\label{checker}} 
\end{figure}

A checkerboard lattice contains 
two spins in the primitive unit cell. 
The corresponding Bravais lattice is formed by
elementary translations on ${\bf a}_1=(1,1)$ and 
${\bf a}_2 =(-1,1)$ and has a square shape. 
In the saturated phase at high fields,
the one-magnon spectrum has, accordingly, 
two branches:
\begin{equation}
\omega_{1{\bf k}}=H-8JS \ ,\ \ \ \ \ \omega_{2{\bf k}}=H - 
4JS(1-\gamma_{\bf k}) \ ,
\label{wchek}
\end{equation}
where $\gamma_{\bf k} = \cos k_x \cos k_y$.
The saturation field $H_s=8JS$ corresponds to the vanishing
energy of magnons from 
the lowest dispersionless branch $\omega_{1{\bf k}}$. 
Localization of excitations
from the lowest branch is determined by the lattice topology.
A simple localized state can be constructed as a 
spin-flip trapped on a square void of a checkerboard
lattice:
\begin{equation}
|\varphi_i\rangle =  \frac{1}{\sqrt{8S}} 
\sum_{n=1}^4 (-1)^{n-1}S^-_{ni}|0\rangle \ ,
\label{localC}
\end{equation}
where the numbering of sites goes counterclockwise starting
from the lowest right corner, see Fig.~\ref{checker}.
The probability to find spin-flip on a site adjacent 
to the void 
vanishes due to the destructive interference.
This property crucially depends on equal strength of
all bonds of the checkerboard lattice.

A linear combination of two square-void states sharing one
vertex encompasses two square voids and so on, see Fig.~\ref{checker}. 
An arbitrary 1-LM state can be constructed by drawing
a closed graph on the original lattice, which passes through 
two vertices of every crossed four-site block,  
and assigning $+$ and $-$£ signs in alternate order
for a spin-flip propagating around it.
The smallest square-void states play 
an important role by forming a complete nonorthogonal
basis in the subspace of dispersionless one-magnon
states.\cite{hard2}
The multi-particle LM states are, then, constructed by
respecting topology of the exchange bonds: 
LMs cannot occupy square voids, which are contiguous to the same
crossed square as shown in Fig.~\ref{checker} by shaded area. 

There is an upper limit on the density of isolated localized
magnons. For a checkerboard lattice, a close-packed structure
is constructed by putting LMs on every second square void 
in horizontal rows with an alternating shift between the rows,
see the left part in Fig.~\ref{checker}. 
This pattern corresponds
to a magnon crystal, which breaks translational symmetry
and has the density of LMs equal to $n_0=N_{\rm max}/N=1/8$.
The breaking of the translational symmetry is, however, 
incomplete: {\it diagonal} rows of LMs can freely slide  
without affecting magnons in adjacent rows.
The degeneracy of the magnon crystal state is
$N_{\rm deg}\sim 2^{L+1}$, where $L$ is a linear size of the system,
see also Sec.~IIc.
Note, that the magnon crystal for a quantum kagom\'e
antiferromagnet is only three-fold degenerate.
In that case there is a well-defined finite temperature
transition associated with the translational
symmetry breaking. 

The purpose of the present study is to investigate the 
low-temperature behavior of the checkerboard antiferromagnet
and nature of a phase transition into the magnon crystal 
state. In order to proceed, we map a quantum checkerboard 
antiferromagnet in the vicinity of the saturation onto a gas of 
hard-core particles. Such an effective model is defined on 
a $\sqrt{2}\times\sqrt{2}$ square lattice formed
by centers of square voids of the original checkerboard lattice.
The dual square lattice is rotated by $45^\circ$ and contains $N/2$ sites.
Localized magnons are represented by hard-core classical particles 
obeying the nearest- and the next-nearest neighbor exclusion principle.
Below, we call them `hard-polygon states.' One such 
polygon for a checkerboard lattice is shown in Fig.~\ref{checker} 
by a shaded square. In reality a state with two localized magnons 
occupying adjacent sites is separated by only a finite gap from 
the low-lying LM states. An estimate of such gap for a kagom\'e
lattice antiferromagnet is given in Ref.~\onlinecite{hard1}.
In the following only $T\rightarrow 0$ regime is considered, where 
contribution of such higher energy states can be neglected.

\subsection{Topological classes of localized magnons}

\begin{figure}[t]
\begin{center}
\includegraphics[width=0.95\columnwidth]{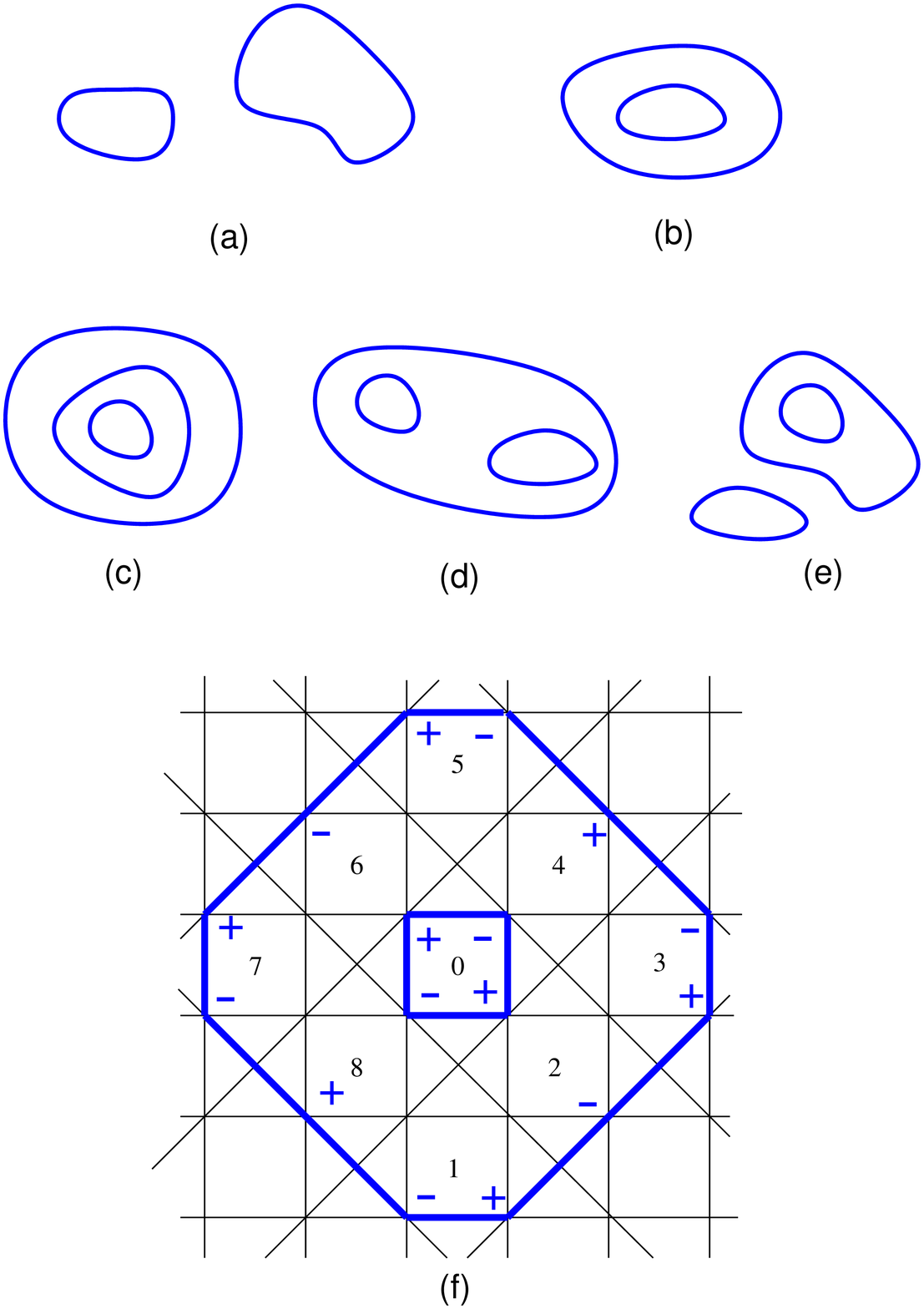}
\end{center}
\caption{(color online). 
Examples of topologically different states of localized magnons:
2-magnon subsector (a) and (b); 3-magnon subsector (c)--(e);
(f) shows an elementary defect state from the class (b).
\label{defect}} 
\end{figure}

The hard-polygon states being linearly independent \cite{schmidt06}
do not exhaust all possible
localized magnon states. The missing `defect' states belong to
different topological classes of LMs. \cite{hard1,hard2}
Let us consider an infinite 
plane with open boundary conditions, which has topology of a sphere 
without the North pole. As was explained before, one-particle 
localized states correspond to closed lines. Successive expansion 
of the wave-function of a one-magnon state in terms of basis plaquette 
states can be represented as a gradual  deformation of a long loop with 
subsequent contraction into a point (center of the last plaquette).
Two closed loops for a 2-LM state may be either contractible
to two distinct points on a plane or lie inside each other 
and, hence, be contractible into a single point, 
see Figs.~\ref{defect}a and \ref{defect}b.
In the first case the two magnon state belongs to a subset 
of two-particle hard-polygon states, whereas in the second case
an expansion in the hard-polygon states is impossible. 
The two-magnon states shown in Figs.~\ref{defect}a and \ref{defect}b
are, therefore, {\it linearly independent}.

Situation is somewhat different for a cluster with periodic
boundary conditions, which has a torus topology. 
In that case one can continuously deform the outer line 
in Fig.~\ref{defect}b, split it into two loops with finite winding 
around the torus, move them around and close 
again into one contour on the opposite side 
such that the two-magnon state in Fig.~\ref{defect}b
transforms into a state of type Fig.~\ref{defect}a.
The linear dependence of two-magnon states in Figs.~\ref{defect}a and 
\ref{defect}b for the torus topology is explained by presence 
in this case of one linear relation between 1-LM square-void (hexagon)
states. \cite{hard2} Still, there are additional 2-LM states 
in this topology, which are constructed
by putting one or two LMs  on contours with nontrivial winding around 
the torus. Topologically nontrivial classes of 
LM states in the three-magnon subsector for a lattice with open 
boundary conditions are shown in Figs.~\ref{defect}c--e. 
For periodic boundary conditions the state in Fig.~\ref{defect}c 
becomes topologically equivalent to the state in Fig.~\ref{defect}e, 
while the state Fig.~\ref{defect}d  can be transformed into 
the hard-polygon state.
The macroscopic limit, however, does not depend on the boundary
conditions.  Equivalence of the two approaches in 
the $N\rightarrow\infty$ limit
is recovered by observing that the largest number of topologically nontrivial
LM states in $n$-magnon subsectors with $n>2$ is given by states of 
the type shown in Fig.~\ref{defect}e, 
which are present for both choices of boundary conditions.
Disentanglement of two, three, etc.\ enclosed loops in the torus topology
is impossible once other LMs are present. On the other hand,
the contribution from closed loops with 
a nontrivial winding around the torus 
corresponds in the thermodynamic limit only to a surface effect.

Topological origin of the additional localized magnon 
states determines their presence for all two-dimensional 
frustrated models including 
antiferromagnets on kagom\'e, checkerboard, and star lattices. 
There are no such states in one-dimensional models
as, for example, the sawtooth chain,
where the hard-particle representation is asymptotically
exact. \cite{hard1,hard2}
In order to estimate the contribution of additional
LMs, one has to define the basis
states in the topologically nontrivial classes.
The elementary defect state in the topological
class of Fig.~\ref{defect}b is shown in Fig.~\ref{defect}f
(open boundary conditions are assumed). Its wave-function is given by 
\begin{equation}
|\textrm{2-defect}\rangle \simeq
\Bigl(\sum_{i=1}^{8} (-1)^{i-1}|\varphi_i\rangle + |\varphi_0\rangle\Bigr) 
|\varphi_0\rangle\ ,
\label{2defect}
\end{equation}
where the numbering of plaquette states follows Fig.~\ref{defect}f.
This wave-function includes two LMs on adjacent voids and on 
the central plaquette and violates, therefore, the hard-polygon 
constraint. The two-magnon states (\ref{2defect}) residing on 
different central squares are linearly independent and form a 
nonorthogonal basis in the subspace of topologically nontrivial graphs 
of Fig.~\ref{defect}b. The 2-LM defect state (\ref{2defect}) 
can be identified with a new classical particle, which has energy
$2\varepsilon_0$ and a longer-range repulsion.
Such a mapping suggests that the topologically nontrivial states
yield an additional macroscopic contribution to the partition
function $\cal Z$.  Their share is, however,
suppressed compared to the hard-polygon contribution by a large
entropic factor: region  
occupied by the basis two-magnon state, Fig.~\ref{defect}f,
can be occupied by 16 different 2-LM states residing
on small empty squares. In addition, by counting the total number of excluded
square voids we conclude that 
the states  (\ref{2defect}) do not contribute appreciably to
the magnetization subsectors with the average density of LMs
larger than $n\agt 2/(25\cdot 2)=0.04$. \cite{hard2} 
Consequently, their role at high densities, {\it e.g.}, in the
vicinity of the transition into a magnon crystal state is strongly
suppressed. Alternatively, the lattice gas mapping can be improved 
by including  extra types of particles,
which describe LMs from different topological classes.
Below we consider the simplest version of the lattice gas mapping 
with only one type of particles.

\subsection{Effective lattice-gas model}

Two-dimensional lattice gas models were originally suggested
to study atomic adsorption on various substrates.
\cite{runnels,ree,bellemans,domany,binder80,kinzel81,schick,kaski,baxter98}
They also describe low-$T$ behavior of Ising antiferromagnets 
in a longitudinal field.\cite{binder80,metcalf,racz}
Despite the long-lasting interest there are only few 
well-established facts for a square lattice 
gas with the nearest- and the next-nearest-neighbor exclusion.
It is generally accepted that this model exhibits a continuous phase
transition into a partially ordered ($2\times 1$)
phase. \cite{ree,kinzel81} 
This phase transition belongs to the 
universality class of an $XY$ model with a four-fold 
anisotropy, \cite{domany} which is a special case of
a more general $Z(4)$ discrete planar model. \cite{elitzur,rujan,wu} 
Depending on the values of two coupling constants
the $Z(4)$ model exhibits either two Ising-like transitions
between a paramagnetic and an ordered state or a single
critical point,\cite{rujan} which has nonuniversal
critical exponents.\cite{jose} 
Estimates for the transition point of a hard-square lattice 
gas with the second-neighbor repulsion vary
from $\ln z_c = (\mu/T)_c=5.3$ in the early work\cite{ree}
to $\ln z_c = 4.7$ in the later study.\cite{kinzel81} 
There is even less certainty about values of the critical exponents.
The entropy of the system at $\mu=0$, which is an interesting
property of a quantum antiferromagnet, has not been
investigated in the context of previous applications.

First, we investigate numerically the entropy of
the lattice gas model at zero chemical potential.
This can be done by the transfer matrix method and
Monte Carlo simulations. The former method 
has been a standard technique since the early studies 
of two-dimensional lattice gas models.\cite{ree,metcalf,kinzel81}
In this scheme a two dimensional lattice
is represented by a semi-infinite strip of width $M$ 
and the thermodynamic properties are
derived from the largest eigenvalue $\lambda_1$ of the transfer matrix.
In particular, the entropy normalized per one site is equal to 
${\cal S} = (\ln\lambda_1)/M$.
The actual calculations become quite simple for $\mu=0$,
when the elements of the transfer matrix take only values 
0 or 1. 
The results for a few values of $M$ are presented in table I.
The convergence is very rapid and already the $M=10$ strip gives
four significant digits for the entropy.

\begin{table}[t]
\caption{The entropy per site $\cal S$ and the density $n$ of 
a hard-square lattice gas with second-neighbor exclusion
at $\mu=0$ obtained from the transfer matrix calculation
on a semi-infinite strip $M\times\infty$.
}
\vspace*{2mm}
\begin{ruledtabular}
\begin{tabular}{rcc}
$M$   &  $\displaystyle {\cal S} = \frac{\ln\lambda_1}{M}$
 &  $n$
\\[3mm]
 \hline
$8$   & $0.294795$ & $0.13713$ \\
$10$  & $0.294671$ & $0.13686$ \\
$12$  & $0.294647$ & $0.13679$ \\
$14$  & $0.294642$ & $0.13677$ 
\end{tabular}
\end{ruledtabular}
\end{table}

To determine entropy from Monte Carlo (MC) simulations 
we adopt the following procedure.
The standard Metropolis algorithm is used for gradual annealing
from the low-activity regime $\ln z=-20$, 
where the density of particles is vanishingly small and ${\cal S}=0$, 
to the point $\ln z=0$. The step for the chemical potential
is chosen to be small enough $\Delta(\ln z)=0.05$. At every value 
of $\mu$  $10^4$ MC steps (lattice sweeps) are performed for equilibration and 
after that $10^5$ MC steps are used to measure 
$(\partial{\cal S}/\partial\mu)_T$, which is calculated 
from the cumulant of the energy $E$ and the number of particles $\cal N$:
\begin{equation}
\left(\frac{\partial {\cal S}}{\partial \mu}\right)_T = 
\frac{1}{T^2}(\langle E{\cal N}\rangle - \langle E\rangle \langle {\cal N}
\rangle) \ .
\label{dSdm}
\end{equation}
In our case $E=-\mu {\cal N}$  and the cumulant in Eq.~(\ref{dSdm})
is proportional to the variance of the total number of particles.
The error bars are estimated by performing up
to 100 independent runs with different sequences of random numbers.
Afterwords, the data for $(\partial{\cal S}/\partial\mu)_T$ are numerically 
integrated to find ${\cal S}|_{\mu=0}$. Simulations have been  
performed on square clusters with periodic boundary conditions
and $N=L^2$ sites, $L=16$, 32, and 64. The obtained results agree 
with each other within numerical accuracy and yield ${\cal S}=0.2946(1)$ 
for the entropy per site and $n=0.13676(1)$
for the average density of particles. The found values are in good 
correspondence with the transfer-matrix results included in table I 
confirming the  
accuracy of both methods. The particle density at $\mu=0$ is 
quite substantial and only two times smaller than the density
of the close-packed structure $n_0=0.25$.

\begin{figure}[t]
\begin{center}
\includegraphics[width=0.8\columnwidth]{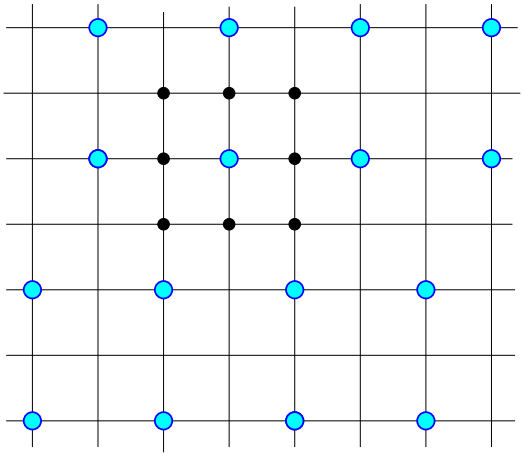}
\end{center}
\caption{(color online). Effective lattice gas model for
a checkerboard antiferromagnet. The dual square lattice is 
$45^\circ$ rotated compared to the original checkerboard lattice 
in Fig.~2 and has twice less sites. Hard particles are shown by 
large light circles; small dark circles indicate the excluded sites
around one particle.
The presented close-packed structure with half-filled rows randomly 
shifted in the horizontal direction has a nonzero Fourier harmonic 
at ${\bf Q}_1=(0,\pi)$.
}
\label{magnonC} 
\end{figure}
 
Once the chemical potential further increases ($z>1$) more and more
particles become condensed. Eventually the lattice gas transforms  
into an ordered close-packed structure.  The order parameter of 
such a crystalline state are certain
Fourier harmonic(s) of the particle density,
$n_{\bf q}=(1/N) \sum_i n_i e^{-i{\bf q}{\bf r}_i}$.
A close-packed structure with half-filled rows, which are randomly 
shifted  in the $x$-direction has a nonzero harmonic at 
${\bf Q}_1=(0,\pi)$ with $n_{{\bf Q}_1}=1/4$,
see Fig.~\ref{magnonC}.
A similar state with random shifts of columns along the $y$ direction
is described by ${\bf Q}_2=(\pi,0)$ and $n_{{\bf Q}_2}=1/4$.
These two phases are called $(2\times 1)$ states in a standard
nomenclature adopted for lattice gas models.\cite{schick} 
A fully symmetric structure with all rows (columns)
in phase with each other has equal amplitudes $n_{{\bf Q}_i}=1/4$
for the three wave-vectors ${\bf Q}_1$, ${\bf Q}_2$, and  
${\bf Q}_3=(\pi,\pi)$ and is called a $(2\times 2)$ state.
The wave-vectors ${\bf Q}_1$ and ${\bf Q}_2$ belong to 
the same irreducible representation. Consequently, we consider 
two order parameters:
\begin{equation}
M_1 = (n_{{\bf Q}_1}^2 + n_{{\bf Q}_2}^2)^{1/2} \ , \ \ \ 
M_2 = |n_{{\bf Q}_3}| \   .
\label{op12}
\end{equation} 

The infinitely degenerate close-packed structure for a 
square-lattice gas with the second-neighbor exclusion
leads to equilibration problems in simple
MC simulations. This was evident since the early MC study
of a frustrated Ising antiferromagnet, \cite{binder80}
which investigated clusters up to only $40^2$ sites 
using a single spin-flip relaxation method.
We employ instead an exchange MC algorithm proposed some time ago
to tackle systems with extremely long relaxation times, as, for example,
spin glasses.\cite{exchange}  
With this modification we have been able to study clusters with up
to $120^2$ sites.
To ensure a substantial replica exchange rate $\sim0.7$--0.9,
we have used 
120 replicas in the range $0\leq\ln z\leq 8$
for all system sizes. 
The simulation runs included $10^5$ exchange MC steps for equilibration and
up to $10^7$ MC steps for measurements.
Statistical errors were estimated from bining the MC series
for each value of $z$.

\begin{figure}
\begin{center}
\includegraphics[width=0.9\columnwidth]{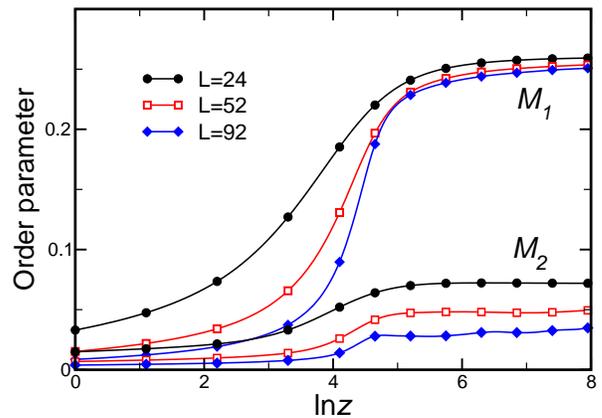}
\end{center}
\caption{(color online). Dependence of the two order parameters
Eq.~(\ref{op12})
on logarithm of the fugacity $\ln z=\mu/T$ for three system sizes.
}
\label{op} 
\end{figure}

\begin{figure}[t]
\begin{center}
\includegraphics[width=0.85\columnwidth]{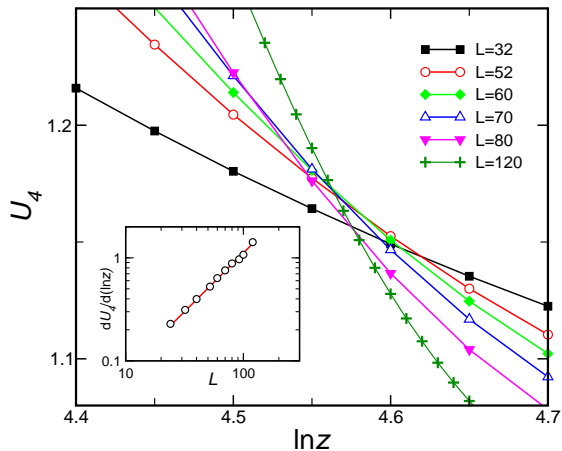}
\end{center}
\caption{(color online). Binder cumulant $U_4$ as a function
of $\ln z=\mu/T$ for different cluster sizes $L$. The inset shows
scaling of the derivative $dU_4/d(\ln z)$ at the transition point
$\ln z_c =4.56$ versus $L$.
}
\label{binder} 
\end{figure}

The specific heat exhibits a broad and rounded maximum near to
$\ln z\sim 4.4$. 
The height of the peak grows very slowly with increasing $L$,
such that a preliminary determination of the transition point
is impossible from the specific heat data. 
The ensemble averages of the two order parameters 
$\langle M_1\rangle$ and $\langle M_2\rangle$ are shown in
Fig.~\ref{op}.
The square of the second order parameter 
$\langle M_2^2\rangle$ goes down to zero as 
$1/N$ with the system size, whereas $\langle M_1\rangle$ 
scales to a finite constant at $\ln z \geq 5$. The limiting value
for the largest cluster is very close $1/4$ in accordance
with the previous analysis for the $(2\times 1)$ type of ordering.
The precise location of the transition point is obtained
by measuring the Binder cumulant 
$U_4=\langle M_1^4\rangle/\langle M_1^2\rangle^2$.
Its dependence on $z$ is presented in Fig.~\ref{binder}.
The transition point can be estimated from the crossing points of 
Binder cumulants for different clusters.  \cite{binder81}
This yields $\ln z_c = 4.56\pm 0.02$ for the critical activity.
The density of particles at the transition point is $n_c=0.2325(3)$.

In principle, the measurement of $M_1 \neq 0$ cannot discriminate
between a single-$k$ structure corresponding to the $(2\times 1)$ 
order and a double-$k$ structure with two nonzero Fourier
harmonics $n_{{\bf Q}_1},n_{{\bf Q}_2}\neq 0$.
The two structures can be distinguished with the help of 
an order parameter, which probes breaking of the $C_4$ 
rotational symmetry:
\begin{equation}
\Delta = \frac{1}{N} \sum_i n_i(n_{i+2{\bf x}}-n_{i+2{\bf y}})  \ .
\end{equation}
The order parameter $\Delta$ vanishes for the $(2\times 2)$ state and 
all other states symmetric under $90^\circ$ rotations, whereas it has 
a finite value in the $(2\times 1)$ phase. We have calculated 
$\langle|\Delta(z)|\rangle$ by MC simulations and found that 
at $\ln z>5$ within statistical errors it reaches a finite value: 
$\langle|\Delta|\rangle=1/8$. This not only demonstrates an absence 
of the tetragonal symmetry in the ordered
state, but also proves {\it random} shifts of half-filled rows (columns).
For example, if shifts occur in a regular alternating
order, than one would find $\langle|\Delta|\rangle=1/4$,
which is definitely excluded by our Monte Carlo results.

In order to determine the critical exponents, we performed
finite-size scaling analysis of various thermodynamic
quantities measured at the estimated critical point $\ln z_c$.
The correlation length exponent $\nu$ is extracted from
the behavior of the derivative of the Binder cumulant
$dU_4/d(\ln z) \sim L^{1/\nu}$ at $z=z_c$. The fitting shown in the inset
of Fig.~\ref{binder} yields $1/\nu = 1.16\pm 0.02$. The order parameter
at criticality scales as $M(z_c)\sim L^{-\beta/\mu}$.
A fit for a few largest $L$ gives $\beta/\nu=0.15(1)$.
As a result, we obtain the following estimates $\nu=0.86(2)$
and $\beta=0.13(1)$. The largest clusters employed
in the present study are still not sufficient
to independently extract the critical exponent $\alpha$,
presumably because of a large regular contribution 
to the specific heat compared to a universal singular part. 
The obtained values for $\beta$ and $\nu$ place a hard-square gas
with the second-neighbor exclusion on a line of critical
points of the $Z(4)$ model between the Potts model, which has
$\beta=1/12$ and $\nu=2/3$,\cite{wu} and the vector
Potts model, which belongs to the Ising universality class 
with $\beta=1/8$ and $\nu=1$. \cite{elitzur} 
Our result for $\ln z_c$ coincides within the error bars
with the corresponding value from the recent independent 
MC study of the same model, \cite{fernandes} though we do 
not share its claim of the Ising universality class 
for the transition.

Properties of a quantum spin-1/2 checkerboard antiferromagnet
in the vicinity of the saturation field are obtained from the above 
results by a straightforward rescaling
to the number of sites of checkerboard lattice, which is twice 
larger than the number of sites for the lattice gas.
At $H=H_s$, the entropy and the magnetization normalized per one site
have universal $T$-independent values:
\begin{equation}
{\cal S}_s = 0.1473\ ,  \ \ \ \ M_s = 0.4316 \ .
\label{Scheck}
\end{equation}
Note, that the entropy of a checkerboard antiferromagnet is 30\% 
larger than the corresponding result for a kagom\'e antiferromagnet. 
\cite{hard1} The transition field into the magnon crystal phase
is given by
\begin{equation}
H_c(T) = H_s - T\ln z_c = 4J - 4.56T \ . 
\label{Hccheck}
\end{equation}
Let us emphasize again that the entropy found within the 
lattice-gas (hard-polygon) description of LMs is only 
approximate. Localized magnons from different topological 
classes will increase the above value of ${\cal S}_s$.
Such corrections should not be very large in view of a large
size of basis defect states (\ref{defect}),
though it would be interesting to estimate 
$\Delta{\cal S}_s$ from an improved lattice gas mapping,
as discussed in the end of Sec.~IIb,
or from the exact-diagonalization data.
Propagating $n$-magnon states can be separated by rather 
small gaps from $n$-LM states. Therefore,  
Eq.~(\ref{Hccheck}) describes the slope
of the actual transition line $H_c(T)$ in
the frustrated quantum antiferromagnet at $T\rightarrow 0$.

\section{Pyrochlore Antiferromagnet}

\subsection{Localized magnons}

\begin{figure}[t]
\begin{center}
\includegraphics[width=0.9\columnwidth]{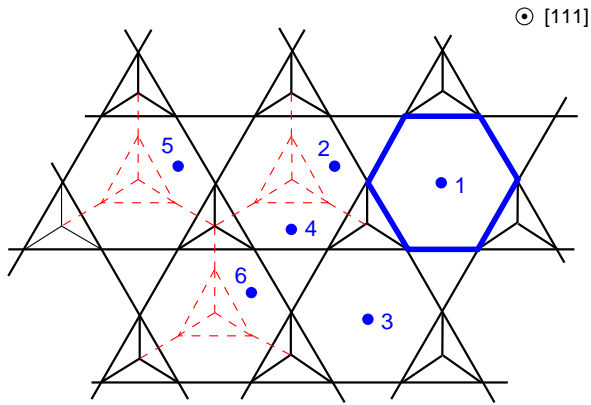}
\end{center}
\caption{(color online). Cross-section of a pyrochlore
lattice perpendicular to the $[111]$ axis.
Tetrahedra from the same kagom\'e layer are drawn by solid lines.
Thick line shows a localized magnon on one of
the hexagons in the kagom\'e plane. Dashed lines indicate
tetrahedra from the upper kagom\'e layer (shown only partially).
Dots with numbers denote centers of hexagons in horizontal and 
tilted planes, which form a dual pyrochlore lattice.
\label{locmag_pyro}} 
\end{figure}

The primitive unit cell of a pyrochlore lattice contains 
4 spins belonging to one tetrahedron. These tetrahedra 
are arranged into an fcc lattice, which is formed by  
the elementary translations on ${\bf a}_1=(0,1/2,1/2)$, 
${\bf a}_2=(1/2,0,1/2)$, and ${\bf a}_3=(1/2,1/2,0)$.
(Linear size of the standard cubic cell with 16 spins is chosen 
as the unit of length. \cite{kennedy}) 
In a strong magnetic field, where the saturated phase 
is stable, the four one-magnon excitation branches 
are given by
\begin{eqnarray}
\omega_{1,2}=H - 8JS,\ \omega_{3,4} = H - 
4JS \pm 2JS \sqrt{1\! +\! \eta_{\bf k}}, && \nonumber \\
\eta_{\bf k}\!=\cos\frac{k_x}{2}\cos\frac{k_y}{2}\!  
+\!\cos\frac{k_y}{2}\cos\frac{k_z}{2}\!+\!\cos\frac{k_z}{2}\cos\frac{k_x}{2}.
&& 
\label{wpyro}
\end{eqnarray}
The saturation field of an antiferromagnet on a pyrochlore lattice
is the same as for a checkerboard lattice: $H_s=8JS$. 
Two dispersionless branches $\omega_{1,2}$ contain 
$2\times(N/4-1)=N/2-2$ linearly independent localized modes, 
$N$ being the number of  pyrochlore lattice sites.
Geometric interpretation of these modes is essentially the same 
as for two-dimensional frustrated lattices.
The smallest LMs are  located on hexagon voids of kagom\'e layers,
which alternate with triangular layers along the $[111]$ and 
three other equivalent directions, see Fig.~\ref{locmag_pyro}.
An $N$-site pyrochlore cluster contains $N$ hexagons, which 
for $N \rightarrow \infty$ is twice more
than the number of localized modes in the two lowest branches 
(\ref{wpyro}). The `elementary' hexagon modes are, therefore, not
only nonorthogonal but also form an overcomplete basis.
This can be explained by observing that hexagon loops on 
a pyrochlore lattice obey exactly $N/2$ linear relations,
which leaves only $N/2$ linearly independent states. \cite{mzh03}
Despite this fact we shall use below the hexagon states 
as a basis in a mapping of a pyrochlore antiferromagnet 
near $H_s$ on a lattice gas model. Such a representation 
gives an incorrect (larger) number of states in the magnetization
subsectors with nonmacroscopic numbers of LMs. 
The mapping becomes not so bad for the multi-magnon subsectors 
in the vicinity of close-packed structures, when the allowed 
configurations do not include linearly dependent states. 
In particular, using the lattice gas representation we shall be 
able to describe an {\it exact} structure of the magnon crystal 
for a pyrochlore antiferromagnet.

The lowest-energy multi-particle states are constructed using 
the same prescription as for two-dimensional frustrated models:
LMs are represented by closed even-sites graphs, which are not 
directly connected by the exchange bonds. 
For example, if a LM occupies a hexagon void 
denoted by 1 in Fig.~\ref{locmag_pyro}, then no LMs can be placed
on hexagons denoted by numbers 2, 3 and 4, since they either share 
sites with the hexagon 1 or are connected to it by the nearest-neighbor
bonds. The lattice gas model is constructed on a dual lattice
formed by centers of hexagons of the original pyrochlore lattice,
such that an exclusion principle for neighboring sites reproduces 
the above rules for LMs.
Remarkably, the dual lattice is again a pyrochlore lattice.
(This explains, in particular, why the number of hexagons
is equal to the number of sites.)
If the nearest-neighbor distance is denoted by $d$
($d\equiv\sqrt{2}/4$), then presence  of a particle on a given site  
excludes for occupation (i) 6 nearest-neighbor sites at the distance $d$ as,
for example, a pair of sites (hexagons) 2 and 4 in Fig.~\ref{locmag_pyro},
(ii) 12 second-neighbor sites at the distance $\sqrt{3}d$, 
{\it e.g.}, a pair of sites 1 and 2, (iii) 6 third-neighbor sites at 
the distance $2d$, {\it e.g.}, sites 1 and 3,  and (iv) 12 fourth-neighbor 
sites at the distance $\sqrt{5}d$, {\it e.g.}, sites 1 and 4.
Note, that on a pyrochlore lattice there are two types of third-neighbor 
sites at the distance $2d$. The first type used above corresponds to
pairs on opposite vertices of hexagons. The second type of third 
neighbors  correspond to pairs with a third site in between.
There are again 6 third-neighbor sites of this type
around a given site. In terms of the original lattice 
they correspond to hexagon pairs (2,5) and (2,6) in 
Fig.~\ref{locmag_pyro}, which allow simultaneous occupation by LMs. 
These hexagons belong to two different parallel kagom\'e planes.

\begin{figure}[t]
\begin{center}
\includegraphics[width=0.95\columnwidth]{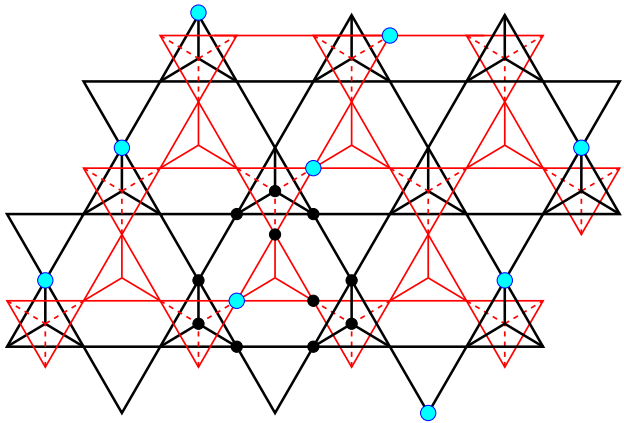}
\end{center}
\caption{(color online). Close-packed structure of particles 
(large light circles) on a dual pyrochlore lattice. 
Thick (thin) lines represent tetrahedra from the lower (upper) 
kagom\'e layer. Small dark circles denote vertices
of one truncated tetrahedron.
}
\label{filling} 
\end{figure}

Let us now consider possible magnon crystal states obtained 
by a close-packing of the smallest LMs on hexagon voids. The 
initial estimate for the maximal density of LMs has been obtained from 
the known result for a single kagom\'e plane: \cite{schulenburg}  
filling one-third of hexagons in every kagom\'e plane, which is  
perpendicular to one of the four cubic diagonals,
gives $n_0 = (1/3)\times(1/4) = 1/12$ 
for the density. No arguments were given yet that this is indeed 
the maximal possible density.
In terms of the effective lattice gas model, which operates
on a dual pyrochlore lattice, a close-packed structure is constructed
by filling every second site in one of the six chains
formed by six edges of one tetrahedron, see Fig.~\ref{filling}. 
Particles at the distance $2d$ along the chains correspond to the second
type of third-neighbors and are allowed by the exclusion principle.
All chains parallel to a chosen direction form a triangular
lattice in the perpendicular plane. Half-filling of the nearest-neighbor
chains is prohibited by the exclusion principle,
whereas second-neighbor chains in the triangular plane
allow for simultaneous occupation.
In this way it is possible to put particles only on one-third of
the parallel chains, which again yields
$n_0= (1/2)\times(1/3)\times(1/2)=1/12$ for the particle
density. (The last factor 1/2 corresponds to the fact that 
parallel chains contain only a half of all lattice sites, the
other half belongs to  the perpendicular chains.) 
An example of the constructed structure is shown in Fig.~\ref{filling}.
Particles in every chain can be independently shifted 
by half a period, which corresponds to the degeneracy $N_{\rm deg}\sim 2^{L^2}$.
Once all chains are in phase, {\it i.e.}, all particles occupy same vertex
of the unit cell tetrahedron, we recover the initial 
structure constructed by filling hexagons in parallel kagom\'e planes.

To prove that the estimate $n_0=1/12$ gives the maximal possible 
density of localized magnons we consider a polyhedron formed by 4 hexagons 
and 4 triangles, which has 12 vertices marked by small circles in 
Fig.~\ref{filling}. This polyhedron is called truncated tetrahedron 
and belongs to the thirteen Archimedean solids.
A special role of this polyhedron for our problem 
is determined by its geometrical 
structure: distance between
two arbitrary vertices of a truncated tetrahedron is equal to $d$, 
$\sqrt{3}d$, $2d$, or $\sqrt{5}d$. If one vertex of a truncated 
tetrahedron is occupied by a particle, then, according to the above
exclusion rules all other vertices must be empty.
Simple consideration yields that an $N$-site pyrochlore
lattice contains $N/2$ truncated tetrahedra such that every 
lattice site is shared between six of them. Counting  
now becomes straightforward: the number of hard-core particles 
in the effective lattice-gas model cannot exceed one particle 
per six truncated tetrahedra $(1/6)\times(N/2)$, which yields 
$1/12$ as the upper bound on $n$. Above, we have constructed 
an explicit example of the particle arrangement with the density $n_0=1/12$,
therefore, this is indeed the highest density of the close-packed
structure.

\subsection{Effective lattice-gas model}

Finite-temperature properties of the effective lattice-gas model
have been studied by Monte Carlo simulations. In the low-density 
disordered regime $z<1$ we use the same annealing protocol 
as in the 2D case and investigate cubic clusters with periodic 
boundary conditions and $N=4L^3$ sites, $L=6$--12. At $\mu=0$ 
($z=1$) the entropy obtained by numerical integration of MC data and 
the density of particles are equal to 
\begin{equation}
{\cal S}/N=0.1329(1)\ , \ \ \ n=0.04718(1)\ ,
\label{Spyro}
\end{equation} 
which yields $M=1/2-n\approx 0.4528$ for the magnetization.

\begin{figure}[b]
\begin{center}
\includegraphics[width=0.9\columnwidth]{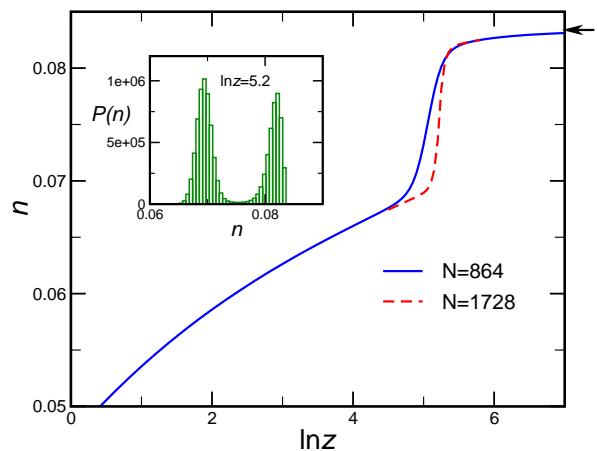}
\end{center}
\caption{(color online). Density of a lattice gas
on a pyrochlore lattice as a function of $\ln z=\mu/T$.
Arrow indicates the density of a close-packed structure
$n_0=1/12$.
Inset: probability distribution for the particle density
obtained for $N=1728$ cluster at $\ln z=5.2$. }
\label{density} 
\end{figure}
 
In order to study the high-density regime $z>1$ we employ 
the exchange MC algorithm with 80 replicas distributed in the
range $0<\ln z<7$. In contrast with the 2D case
we still find significant equilibration problem for large lattices,
which exhibit extremely long relaxation times $t \agt 10^7$ MC steps 
despite of a high exchange rate $\sim 0.8$ between replicas. 
Sufficient statistics has been obtained only for three system sizes:
two cubic clusters with $L=4,6$ and a tetragonal prism with 
$6\times 6\times 12$ unit cells (1728 sites). They were equilibrated 
for $5\times 10^5$ exchange MC steps and further 
$10^7$ MC steps were used for measurements.
Evolution of the particle density with increasing $z$ is presented in
Fig.~\ref{density} for the two biggest systems.
In the vicinity of $\ln z \approx 5.5$ there is an abrupt jump of $n$
to the value, which is very close to the density of the close-packed
structure $n_0=1/12$.

We expect that the ordered state at $\ln z\agt 5.5$ exhibits only a 
partial breaking of the translational symmetry.
Similar to the studied 2D model, the half-filled 
chains remain uncorrelated. This tendency should be even more pronounced
for the pyrochlore lattice gas in view of an increased entropic effect 
from disordering determined by a larger $N_{\rm deg}$. 
As a result, the symmetry breaking 
includes only (i) selection of a chain direction and (ii)
formation of $\sqrt{3}\times\sqrt{3}$ structure on 
a triangular lattice in the perpendicular plane. 
In view of small available system sizes we have not performed
investigation of the corresponding order parameters.
Instead we have tried to clarify 
the nature of a possible phase transition into a close-packed
structure by collecting histograms for the particle density. 
Results for $N=1728$ lattice at $\ln z=5.2$ obtained
with $10^7$ MC steps are presented in the inset 
of Fig.~\ref{density}.
A clear double peak structure with approximately equal weights 
in each peak points at a first-order transition.
This should be, of course, confirmed by a similar study of larger clusters.
We, therefore, put a tentative estimate $\ln z_c=5.2\pm 0.2$ 
for the first-order transition into the magnon crystal state for 
a pyrochlore antiferromagnet. 

Since the original and the dual pyrochlore lattices have the same
number of sites, the properties of a spin-1/2
pyrochlore antiferromagnet are obtained from the above results
without any rescaling. In particular, the estimate for the
entropy at the saturation field is given by Eq.~(\ref{Spyro}).
At zero temperature and $H=H_s$ the magnetization jumps from  
$M=5/12$ in the plateau region to the full saturation with $M=1/2$.
At finite temperatures the jump in the magnetization is much smaller
$\Delta M \approx 0.015$, see Fig.~\ref{density}.
Afterward, $M(H)$ increases smoothly and  asymptotically 
saturates. The first-order transition into the magnon crystal state
takes place at 
\begin{equation}
H_c(T) = H_s -T\ln z_c\approx 4J - 5.2 T 
\end{equation} 
for a spin-1/2 model.

\section{Effect of extra exchanges}

In the previous two sections, we have found that the magnon crystal states 
in checkerboard and pyrochlore antiferromagnets exhibit massive degeneracy
at $T=0$. The degeneracy is not lifted
at finite temperatures by fluctuations within the LM subspace. This leads 
to interesting types of broken symmetries in 2D and 3D cases. In principle, 
fluctuations to higher energy states outside the LM ensemble at $T>0$ can favor one 
or the other type of close-packed magnon structures. Such a scenario 
cannot be excluded on general grounds. The higher energy 
states are, however, separated by finite gaps from LMs. Therefore,
there must be a region near $T\rightarrow 0$, $H\rightarrow H_s-0$, where 
such a selection is ineffective and only a partial breaking 
of translational symmetry takes place. Another mechanism of degeneracy 
lifting can be provided by the magnetoelastic coupling, \cite{richter04} 
though the corresponding study has considered only one possible 
pattern of LMs for each of the two lattices.
In the following we investigate the third possibility: effect of further 
neighbor exchanges, \cite{reimers} which play significant role in
some magnetic materials with pyrochlore lattices.
\cite{crawford,tsune,cepas,motome,wills}

We begin with the checkerboard antiferromagnet, considering an additional
weak diagonal exchange of strength $J'$, see Fig.~\ref{lattices}b.
The magnon crystal state becomes completely unstable at a certain critical
value of $J'$. \cite{hard2}
We are interested in the subcritical regime $|J'|\rightarrow 0$.  
Let us consider two localized magnons (\ref{localC}) on adjacent square 
plaquettes along the $[11]$ direction: 
$|\varphi_i\rangle$ and $|\varphi_{i+2{\bf a}_1}\rangle$, see 
Fig.~\ref{checker} and  Sec.~IIa for the notations. These LMs are connected by  
an additional exchange bond  
$\hat{V}=J'{\bf S}_{2,i}\cdot{\bf S}_{4,i+2{\bf a}_1}$.
To measure energy shifts relative to the energy
of the fully polarized state, we shall always subtract 
a constant from the exchange bond operators: $\hat{V} \rightarrow \hat{V} - J'S^2$. 
The potential energy $U$ of two LMs on adjacent voids is 
found by calculating the 
expectation value of $\hat{V}$ over the two-magnon state 
$|\psi\rangle=|\varphi_i\varphi_{i+2{\bf a}_1}\rangle$ 
and correcting it by the self-energy of LMs without neighbors:
\begin{equation}
U = \langle\psi|\hat{V}|\psi\rangle - 
2 \langle\varphi_i|\hat{V}|\varphi_i\rangle  
=  J'/16 \ .
\label{Echecker}
\end{equation}
This result applies to an arbitrary spin value $S$.
For $J'>0$ presence of LMs on adjacent square voids 
is energetically unfavorable, because 
two spin flips  with a finite probability 
occupy the same $J'$ bond. Degeneracy of the magnon crystal is, 
consequently, lifted in favor of an orthorhombic structure:
half-filled rows of LMs alternate in phase in order
to minimize the contribution (\ref{Echecker}) between the rows.
For $J'<0$  localized magnons attract each other 
and the $(2\times 2)$ tetragonal close-packed structure is stabilized. 

The interaction energy between two LMs residing on hexagon voids of 
pyrochlore lattice is obtained in the same way as Eq.~(\ref{Echecker}) 
with the result $U=J'/36$. For a pyrochlore antiferromagnet 
three additional exchanges may be present:\cite{reimers,wills} 
the second-neighbor exchange $J_2$ 
for spin pairs at distance
$\sqrt{3}d$, the third-neighbor exchange $J_{31}$ for spin
pairs at distance $2d$ on opposite vertices of hexagons,
and the third-neighbor exchange $J_{32}$ for spin pairs
at distance $2d$ along chains, see Fig.~\ref{lattices}a. 
(Details about the crystal structure of pyrochlore oxides
can be found, {\it e.g.}, in Ref.~\onlinecite{kennedy}.)
Similar to the 2D case the weak additional exchanges 
select relative shift between adjacent half-filled  chains of LMs.
If two chains are not shifted relative to each other,
LMs occupy hexagons in parallel kagom\'e planes,
whereas if the shift is present, LMs form a nonplanar structure.
Analysis of further neighbor exchange links between hexagons 
on a pyrochlore lattice is straightforward but cumbersome.
Below we only summarize the conclusions.
Once  the two chains of LMs form a planar structure,
one LM from the first chain is coupled to two
hexagons in the second chain: to a hexagon in the same kagom\'e plane
by two $J_2$ and two $J_{32}$ bonds and to a hexagon 
in a parallel  kagom\'e plane by a $J_{31}$ bond.
Then, the interaction energy between chains
normalized per one LM of the first chain is 
$U_1 = (2J_2+J_{31}+2J_{32})/36$. 
In the nonplanar structure 
a given LM from the first chain is coupled to only one
non-planar hexagon by three $J_2$ bonds and one $J_{32}$ bond,
which yields $U_2=(3J_2+J_{32})/36$.

Comparing the interaction energies $U_1$ and $U_2$ for the two
structures, we conclude that for $(J_{31}+J_{32}-J_2)<0$  
localized magnons occupy hexagons in parallel kagom\'e planes. Between
the planes LMs follow the ABCABC... structure
of close-packed hard spheres.
If the above expression changes sign, then LMs form a more 
complicated nonplanar structure. Amazingly enough, it is again
frustrated and degenerate: half-filled chains form a triangular lattice,
therefore, antiphase shifts would correspond to an effective Ising
antiferromagnet on a triangular lattice.
The residual degeneracy should be finally lifted by zero-point fluctuations 
induced by the next-neighbor exchanges. The corresponding energy 
scale is of the order of $J'^2/J$.

\section{Summary}

Knowledge of the exact ground state and the single-particle excitation
spectrum of frustrated quantum antiferromagnets above the saturation
field $H_s$ allows to develop a quantitative description 
of their thermodynamic properties in a finite range of 
fields $H\alt H_s$ and temperatures $T\rightarrow 0$. 
Localized nature of the lowest energy excitations leads to a mapping
onto lattice gas models of hard-core classical particles with 
an appropriate exclusion principle. 
As a result, frustrated quantum antiferromagnets preserve
a macroscopic degeneracy (finite entropy) at $H=H_s$ and $T=0$.
The close-packed structure of 
particles (localized magnons) corresponds to the magnon crystal state,
which breaks only translational symmetry and preserves
continuous rotations about the field direction.

The main finding of our study is high degeneracy of 
the magnon crystal states in pyrochlore and checkerboard
antiferromagnets at zero temperature.
This, however, does not exclude  presence of finite-temperature
phase transitions. 
The corresponding lattice gas models have been studied by the 
exchange Monte Carlo method. 
Specifically, we have found numerically the location
$\ln z_c = 4.56(2)$ and two critical exponents $\beta = 0.13(1)$
and $\nu=0.86(2)$ for a square-lattice gas with the second-neighbor
exclusion, which describes a checkerboard antiferromagnet. 
Further numerical work is needed to clarify small differences 
with the recent independent MC study of the same model.\cite{fernandes}
Monte Carlo simulations for the pyrochlore-lattice gas show
a sharp jump in the particle density, which suggests a first-order
transition into a partially ordered crystal phase.

\acknowledgments

We are grateful to L. Balents and A. Honecker for useful discussions.
We also thank to H.-J. Schmidt for a helpful remark
on topological classes of localized magnons.
This work was partly supported by a Grant-in-Aid
for Scientific Research on Priority Areas (No.~17071011)
and Scientific Research (No.~16540313), and also by
the Next Generation Super Computing Project,  Nanoscience
Program, from the Ministry of Education, Culture,
Sports, Science and Technology of Japan.


\begin{thebibliography}{99}

\bibitem{villain}
J. Villain, R. Bidaux, J. P. Carton, and R. J. Conte, 
J. de Phys. {\bf 41}, 1263 (1980).

\bibitem{shender}
E. F. Shender, 
Sov. Phys. JETP {\bf 56}, 178 (1982).

\bibitem{henley89}
C. L. Henley,
Phys. Rev. Lett. {\bf 62} 2056 (1989).

\bibitem{RK}
D. S. Rokhsar and S. A. Kivelson, 
Phys. Rev. Lett. {\bf 61}, 2376 (1988).

\bibitem{cabra}
D. C. Cabra, M. D. Grynberg, P. C. W. Holdsworth, A. Honecker, 
P. Pujol, J. Richter, D. Schmalfuss, and J. Schulenburg,
Phys. Rev. B {\bf 71}, 144420 (2005)

\bibitem{mzh05}
M. E. Zhitomirsky,
Phys. Rev. B {\bf 71}, 214413 (2005).

\bibitem{bergman}
D. L. Bergman, R. Shindou, G. A. Fiete, and L. Balents,
Phys. Rev. Lett. {\bf 96}, 097207 (2006).

\bibitem{hizi}
U. Hizi and C. L. Henley,
Phys. Rev. B {\bf 73}, 054403 (2006).

\bibitem{hard1}
M. E. Zhitomirsky and H. Tsunetsugu,
Phys. Rev. B {\bf 70}, 100403(R) (2004).

\bibitem{hard2}
M. E. Zhitomirsky and H. Tsunetsugu,
Prog. Theor. Phys. Suppl. {\bf 160}, 361 (2005).

\bibitem{derzhko04}
O. Derzhko and J. Richter,
Phys. Rev. B {\bf 70}, 104415 (2004).

\bibitem{richter06}
J. Richter, O. Derzhko, T. Krokhmalskii,
Phys. Rev. B {\bf 74}, 144430 (2006).


\bibitem{baxter}
R. J. Baxter, {\it Exactly Solved Models in Statistical Mechanics}
(Academic Press, London, 1982).

\bibitem{schnack}
J. Schnack, H.-J. Schmidt, J. Richter, and J. Schulenburg,
Eur. Phys. J. B {\bf 24}, 475 (2001).

\bibitem{schulenburg}
J. Schulenburg, A. Honecker, J. Schnack, J. Richter, and H.-J. Schmidt,
Phys. Rev. Lett. {\bf 88}, 167207 (2002).

\bibitem{schmidt}
H.-J. Schmidt, 
J. Phys. A {\bf 35}, 6545 (2002).

\bibitem{mzh03}
M. E. Zhitomirsky, 
Phys. Rev. B {\bf 67}, 104421 (2003).

\bibitem{richter}
J. Richter, J. Schulenburg, A. Honecker, J. Schnack, and H.-J. Schmidt,
J. Phys.:\ Condens.\ Matter {\bf 16}, S779 (2004).

\bibitem{schmidt06}
H.-J. Schmidt, J. Richter, and R. Moessner,
J. Phys. A {\bf 39}, 10673 (2006).

\bibitem{runnels}
L. K. Runnels,
Phys. Rev. Lett. {\bf 15}, 581 (1965).

\bibitem{ree}
F. H. Ree and D. A. Chestnut, 
Phys. Rev. Lett. {\bf 18}, 5 (1967).

\bibitem{bellemans}
A. Bellemans and R. K. Nigam, 
J. Chem. Phys. {\bf 46}, 2922 (1967).

\bibitem{domany}
E. Domany and E. K. Riedel,
Phys. Rev. Lett. {\bf 40}, 561 (1978).

\bibitem{binder80}
K. Binder and D. P. Landau,
Phys. Rev. B {\bf 21}, 1941 (1980).

\bibitem{kinzel81}
W. Kinzel and M. Schick,
Phys. Rev. B {\bf 24}, 324 (1981).

\bibitem{schick}
M. Schick,
Physica B {\bf 109\&110}, 1811 (1982).

\bibitem{kaski}
K. Kaski, W. Kinzel and J. D. Gunton,
Phys. Rev. B {\bf 27}, 6777 (1983).

\bibitem{baxter98}
R. J. Baxter,
Ann. Comb. {\bf 3}, 191 (1999). 

\bibitem{metcalf}
B. D. Metcalf and C. P. Yang, Phys. Rev. B {\bf 18}, 2304 (1978).

\bibitem{racz}
Z. R\`acz, 
Phys. Rev. B {\bf 21}, 4012 (1980).

\bibitem{elitzur}
S. Elitzur, R. B. Pearson, and J. Shigemitsu,
Phys. Rev. D {\bf 19}, 3698 (1979).

\bibitem{rujan}
P. Ruj\'an, G. O. Williams, H. L. Frisch, and G. Forg\'acs,
Phys. Rev. B {\bf 23}, 1362 (1981).

\bibitem{wu}
F. Y. Wu, Rev. Mod. Phys. {\bf 54}, 235 (1982).

\bibitem{jose}
J. V. Jos\'e, L. P. Kadanoff, S. Kirkpatrick, and D. R. Nelson,
Phys. Rev. B {\bf 16}, 1217 (1977).

\bibitem{exchange}
K. Hukushima and K. Nemoto,
J. Phys. Soc. Jpn. {\bf 65}, 1604 (1996).

\bibitem{binder81}
K. Binder,
Z. Phys. B {\bf 43}, 119 (1981).

\bibitem{fernandes}
H. C. M. Fernandes, J. J. Arenzon, and Y. Levin,
{\tt arXiv:cond-mat/0612372v1}.

\bibitem{kennedy}
B. J. Kennedy, B. A. Hunter, and C. J. Howard,
J. Solid State Chem. {\bf 130}, 58 (1997).

\bibitem{richter04}
J. Richter, O. Derzhko, and J. Schulenburg,
Phys. Rev. Lett. {\bf 93}, 107206 (2004);
O. Derzhko and J. Richter, 
Phys. Rev. B {\bf 72}, 094437 (2005).

\bibitem{reimers}
J. N. Reimers, A. J. Berlinsky, and A.-C. Shi,
Phys. Rev. B {\bf 43}, 865 (1991).

\bibitem{crawford}
M. K. Crawford, R. L. Harlow, P. L. Lee, Y. Zhang, J. Hormadaly, R. Flippen, 
Q. Huang, J. W. Lynn, R. Stevens, B. F. Woodfield, J. Boerio-Goates, and R. A. Fisher,
Phys. Rev. B {\bf 68}, 220408(R) (2003). 

\bibitem{tsune}
H. Tsunetsugu and Y. Motome,  
Phys. Rev. B {\bf 68}, 060405(R) (2003). 

\bibitem{cepas}
O. Cepas and B. S. Shastry,
Phys. Rev. B {\bf 69}, 184402 (2004).

\bibitem{motome}
Y. Motome and   H. Tsunetsugu, 
Phys. Rev. B {\bf 70}, 184427 (2004); 
Prog. Theor. Phys. Suppl. {\bf 160}, 203 (2005).

\bibitem{wills}
A. S. Wills, M. E. Zhitomirsky, B. Canals, J. P. Sanchez, 
P. Bonville, P. Dalmas de Reotier, and A. Yaouanc,
J. Phys.: Condens. Matter {\bf 18}, L37 (2006).


\end{thebibliography}
\end{document}